\documentclass[a4paper,10pt]{article}
\usepackage[utf8x]{inputenc}
\usepackage{threeparttable}
\pdfoutput=1
\usepackage{verbatim}
\usepackage{epsfig}
\usepackage{epsf}
\usepackage{graphicx}
\usepackage{textcomp}
 \usepackage{amsmath}
\usepackage{color}
\definecolor{Blue}{rgb}{0.3,0.3,0.9}
\definecolor{red}{rgb}{1,0,0}

\newcommand{\bM}[1]{\mathbf{#1}}    

\newcommand{\vecthree}[3]
{
   \begin{pmatrix} #1 \\ #2 \\ #3\end{pmatrix}
}

\title{\textbf{\fontsize{15}{15}\selectfont Photon Trajectory in the Human Cornea}}
\author{\smallskip \textbf{\fontsize{10}{10}\selectfont Stephen G. Odaibo}\\ \textbf{\fontsize{10}{10}\selectfont M.S.(Math), M.S.(Comp. Sci.), M.D.}\\ \newline\newline \small{\textsl{\fontsize{7}{7}\selectfont Howard University Hospital, Department of Ophthalmology, Towers Bldg. Suite 2100}}\\ \small{\textsl{\fontsize{7}{7}\selectfont 2041 Georgia Avenue N.W.}}\\ \small{\textsl{\fontsize{7}{7}\selectfont Washington, D.C. 20060}}\\ \small{\color{blue}{\fontsize{7}{7}\selectfont photon.biology@gmail.com}}}

\date{}

\begin{document}

\maketitle

\begin{abstract}
In this article, we follow the trajectory of a photon in the human cornea. Prior to experimental evidence of the cornea's gradient refractive index  (GRIN) nature, schematic eye models used a constant or average as the corneal refractive index. A few recent models respect the intra-corneal GRIN, but are based on the paraxial approximation, and thereby have limited validity in peripheral visual field analysis and non-paraxial photon tracking. Here, we introduce the Trajectron algorithm. It uses a closed form solution of the ray equation of constant axial GRIN media, and evaluates over piece-wise constant GRIN. Using Trajectron, we present the first quantitative demonstration of non-paraxial intra-corneal light-bending phenomena. This demonstration has significant implications in the development of refractive surgery algorithms.

\end{abstract}
\hspace{-1.5mm}\indent\indent {{\fontsize{7}{7}\selectfont {\bf Keywords:} Cornea, Schematic eye, GRIN optics, Refractive surgery}}
\begin{center}
\line(1,0){300}
\end{center}

\section{Introduction}
\subsection{Aim}
To determine the trajectory of a photon being observed in the human cornea, from a point within the anterior central corneal epithelium to a point within the posterior stroma.

\subsection{Motivation}
Refractive error places an enormous disease burden on society, with net estimates of over two billion people affected worldwide~\cite{absi2009,ea2004,viel2008,xuli2005,wakl94,fala2004,sa2003,saga2002,dada99,sacha2008,ezra2011}. Myopia, hyperopia, presbyopia, astigmatism, and cataracts are among the most common forms of refractive error. Refractive surgery is an effective and growing method of correcting refractive error. There are a high volume of cases performed annually, including cataract extractions with intraocular lens placement, laser-assisted in-situ keratomileusis (LASIK) by microkeratome incision or by femtosecond laser incision, photorefractive keratectomy (PRK), and laser-assisted sub-epithelial keratomileusis (LASEK). Refractive surgery techniques and technologies have seen significant advances over the years including current reports of greater than 90\% patient satisfaction post-operatively~\cite{tabo2005,bami2003}. However, outcomes remain suboptimal and patients are often still left to manage imperfect vision following surgery. Most modern refractive surgery techniques involve direct modification of the corneal epithelium and stroma. It is therefore important to have a precise and detailed understanding of the refractive properties along this path. It is additionally imperative to understand how those properties govern the behavior of light both as a source of visual stimulus, and in amplified form, as lasers for refractive and other ophthalmic intervention. Furthermore, a precise quantitative mapping of the refractive properties of the human cornea is requisite for good design of keratoprosthetic implants. In this paper, we embark on a mathematical exploration of the photon trajectory from the anterior central corneal epithelium to the posterior stroma. We operate within the realm of geometric optics, which assumes the limit of short wavelength, where light propagation can be approximated by rays and studied using geometry. We introduce the Trajectron algorithm for computing the photon trajectory in a model respectful of the cornea's axial GRIN nature. Meridional intra-corneal GRIN is of significantly less magnitude than axial intra-corneal GRIN, and is assumed to be zero in this model. Trajectron uses a closed form solution of the ray equation of constant axial GRIN media, and evaluates over piece-wise constant GRIN. One advantage of Trajectron is that it is not restricted to the paraxial regime. This makes it eligible for the study of non-paraxial optical phenomena such as peripheral visual field analysis, as well as higher-order aberrations before and after refractive surgery. 

\subsection{Related work}

Previous mathematical models of the human eye assumed the corneal refractive index to be constant~\cite{atsm2000,sm1995,ma57,ma1994}. Such models were proposed when less was known about the intra-corneal GRIN nature. A few recent models based on ray-transfer matrix (ABCD) analysis incorporate the intra-corneal GRIN nature. Barbero proposed a multilayer model of cornea and tear-film to calculate corneal power. His global ray-transfer matrix was a product of 7 matrices representing ray transfer through 4 interfaces and 3 media which included a GRIN stroma~\cite{ba2006}. Flores-Arias et al proposed a linear cascade model expressing the generalized fresnel integral transform kernel in terms of paraxial coefficients; where the global ray-transfer matrix was a product of anterior chamber, GRIN cornea, and GRIN lens component ray-transfer matrices~\cite{fldi2009, peba2005}. These newer models incorporating the GRIN nature represent a significant advance. One disadvantage however, is that they are restricted to the paraxial regime, and this limits their validity for peripheral visual field analysis and non-paraxial photon tracking. In contrast, our current work is useful for investigating interesting non-paraxial optical phenomenology, because it respects the intra-corneal GRIN nature, and is not restricted to the paraxial regime. 

\subsection{Contributions}
In this paper we make the following technical and biophysical contributions:

\begin{itemize}
\item The first quantitative demonstration of non-paraxial light-bending within the human cornea.
\item An analytico-numerical algorithm, Trajectron, for determining the photon trajectory within the human cornea, respecting axial GRIN and assuming zero meridional GRIN.
\item A method for studying non-paraxial optical phenomena in the human cornea. For example:
\begin{itemize}
 \item Peripheral visual field analysis
\item Higher order aberrations pre and post refractive surgery
\end{itemize}
 \item An approach towards describing the angular domain over which the paraxial approximation is clinically useful in the human cornea.
\end{itemize}

\subsection{Methods}
We use a hybrid approach inter-weaving between analytical and numerical processes. Our objective is to determine the ray path from a point in the anterior central corneal epithelium to a point in the posterior stroma. The human cornea has an axial GRIN of non-constant magnitude~\cite{pama95}, and we approximate it as also having zero meridional GRIN. This approximation is made because, although meridional intra-corneal GRIN has been demonstrated in-vivo in humans~\cite{vasi08}, it is significantly less than the axial intra-corneal GRIN~\cite{pama95}. To achieve our objective, we introduce Trajectron, a new algorithm which evaluates a closed form solution of the ray equation of constant axial GRIN media, over piece-wise constant axial GRIN. The details of the Trajectron algorithm are presented in Section~(\ref{sec:rslt}) below. Discrete measurements of corneal refractive indices are obtained from the literature, and a continuous distribution is generated by curve fitting. Trajectron is prototyped in MATLAB{\fontsize{6}{6}\textregistered} (The MathWorks Inc., Natick, MA). 

The remainder of this paper is organized as follows: Section~(\ref{sec:PhysFor}) outlines the physical formalism of lagrangian optics and includes a known closed form solution of the ray equation of constant axial GRIN media; Section~(\ref{sec:Msmt}) describes structure, thickness, and refractive index measurements within the human cornea; Section~(\ref{sec:compu}) introduces the Trajectron algorithm; Section~(\ref{sec:rslt}) presents the results of our numerical experiments; Section~(\ref{sec:disc}) discusses the findings, the study limitations, and the clinical angular domain of the paraxial and non-paraxial regimes; and Section~(\ref{sec:conclu}) concludes the paper. 

\section{Physical Formalism}
\label{sec:PhysFor}
Fermat's principle of stationary time asserts that light travels along the path that takes the least (or greatest) time. In the context of a classical mechanics treatment, Fermat's principle is fully described by the lagrangian of geometric optics, given by,

\begin{equation}
 {\cal{L}} = n(x,y,z)\sqrt{1+{\dot{x}}^2+{\dot{y}}^2},
\label{Optical_Lagrangian}
\end{equation}

where $n$ is the refractive index distribution, $x$, $y$, and $z$ are spatial coordinates, and $\dot{q}:=\frac{\partial q}{\partial z}$. While $x$ and $y$ take the role of generalized coordinates, $z$ describes the optical axis and takes the role of parameter. 

The corresponding Euler-Lagrange equations are given by,

\begin{equation}
\frac{\partial}{\partial z} \left( \frac{\partial {\cal L}}{\partial \dot{q}} \right) =
 \frac{\partial{{\cal L}}}{\partial{q}},
\label{Fermat_E-L}
\end{equation}

where Equation~(\ref{Fermat_E-L}) encodes three equations: with $q=x$, $q=y$, and $q=z$.

Applying Equation~(\ref{Fermat_E-L}) above to the optical lagrangian, Equation~(\ref{Optical_Lagrangian}), yields the ray equation, 

\begin{equation}
\frac{\partial}{\partial{s}}\left(n\frac{\partial\mathbf{r}}{\partial s}\right) = \nabla n,
\label{ray_eqn} 
\end{equation}

which represents the equations of motion for propagation of a light ray through a medium of refractive index distribution $n(x,y,z)$. In Equation~(\ref{ray_eqn}), $\mathbf{r}$ is the spatial position vector $[x, y, z]$ of the ray, $\nabla$ is the gradient operator, and $\partial{s}$ is the arc length along the ray trajectory, and is given by $\partial{s} = \partial{z}\sqrt{1+{\dot{x}}^2 + {\dot{y}}^2}$.

In the case of constant refractive index gradient, $\alpha$, along the optical axis direction, and zero refractive index gradient along the other two cardinal directions, Equation~(\ref{ray_eqn}) can be written as,

\begin{equation}
 \frac{\partial}{\partial{s}}\left(n\frac{\partial\mathbf{r}}{\partial s}\right) = \vecthree{0}{0}{\frac{\partial{n}}{\partial{z}}}.
\label{ray_eqn_const_grad} 
\end{equation}

The ray $\mathbf{r}=[x, y, z]$ is embedded in a 2-dimensional plane. This plane of propagation is defined by the optical axis and $d\bM{s}\big|_i$, where $i$ denotes the point along the optical axis in the anterior corneal epithelium where we initiate observation, and $d\bM{s}\big|_i$ denotes the ray direction at $i$. The embedding, $(x, y, z)\longrightarrow (r,z)$, is given by,

\begin{equation}
\begin{array}{l}
\displaystyle x = r ~\mbox{cos} (\psi),\\
\displaystyle y = r ~\mbox{sin} (\psi),\\
\displaystyle z = z,
\end{array} 
\label{ray_embedding}
\end{equation}

where $\psi$ is the angle the propagation plane makes with the positive $y$-axis.

The components of the canonical optical momentum are given by,

\begin{equation}
 p_1 = n\frac{dx}{ds},\qquad p_2 = n\frac{dy}{ds}, \qquad \mbox{and} \qquad p_3 = n\frac{dz}{ds},
\label{canonical_optical_momenta}
\end{equation}

and as follows for direction cosines, sum of squares equals one, hence,

\begin{equation}
 |\bM{p}|^2= n^2.
\label{momentum_conservation}
\end{equation}

 Equations~(\ref{ray_eqn_const_grad}) and (\ref{canonical_optical_momenta}) together state that $p_1$ and $p_2$ are constant along the ray trajectory. Equation~(\ref{ray_eqn_const_grad}) in conjunction with the constancy of $p_1$ and $p_2$, constrain the ray $\bM{r}$ to the $(r,z)$ plane.

Applying the fundamental theorem of calculus to the conservation relation for optical momentum shown in Equation~(\ref{momentum_conservation}), one obtains the following known closed form solution of the ray equation in constant axial GRIN media~\cite{Qi84},

\begin{equation}
 r(z)=A ~{\ln\left(2\alpha^2 z  + 2{\alpha}G + 2\alpha n_i\right)}\bigg|_i^z,  
\end{equation}

where $n_i$ is the refractive index at the initial observation point of the photon, $G$ is a function of $z$ given by,

\begin{equation}
 G(z) = \sqrt{ \alpha^2z^2 + 2\alpha n_iz + n_i^2{\sin}^2({\phi}_i)},
\label{G_eqn}
\end{equation}

$A$ is a constant given by,

\begin{equation}
 A =\frac{n_i\cos({\phi}_i)}{\alpha},
\label{A_const}
\end{equation}

and ${\phi}_i$ is the angle between $d\bM{s}\big|_i$ and the positive $r$ axis.

The angle between the optical axis and  $d\bM{s}\big|_i$ is denoted by  $\theta_i$. In this paper we are primarily interested in rays entering the cornea, and will therefore focus on photon trajectories for which $0 \leq \theta_i < \frac{\pi}{2}$. 
  
\section{Measurement}
\label{sec:Msmt}

Closed-form and numerical solutions of the ray equation of axial GRIN media require knowledge of the GRIN. The gradients can be approximated by the difference quotients. This requires appropriately sampled measurements of refractive index and thickness. Below, we examine the literature for the required measurements in the four corneal layers along the photon trajectory of interest: (i) corneal epithelium, (ii) Bowman's layer, (iii) anterior stroma, and (iv) posterior stroma.  

\subsection{Corneal epithelium}

The corneal epithelium is a non-keratinized stratified squamous epithelium~\cite{kl77} of 5-7 cell layers with rapid surface turn over. The corneal epithelium is covered by mucinous tear film anteriorly, and is adjacent to  Bowman's layer posteriorly. The morphology of the air-tear film interface approximately conforms to the corneal epithelial surface. Corneal epithelial surface layer abrasions can cause significant visual distortion, and can manifest photophobia and moderate to severe amounts of pain. However as a result of limbal stem cell activity~\cite{lure2001}, abrasions typically heal in 24-48 hours in healthy individuals. Both LASEK and PRK procedures involve manipulation of the corneal epithelium. In PRK, the epithelial layer is removed entirely by mechanical or alcohol debridement, to allow for laser ablation of the anterior stroma~\cite{muko1988,brsh2003}. In LASEK, an alcohol solution is used to disrupt the epithelial layer, creating a flap which is then replaced following laser ablation of anterior stroma~\cite{azan2001,brsh2003}.

\subsubsection{Corneal epithelial thickness}

There have been several measurements of human central corneal epithelial thickness by various modalities. Significant variation exists in the findings, but almost all the results have been in the 49-60 $\mu$m range. The following is a sample of prior work in this area. King-Smith et al used interferometry to make measurements on 6 healthy eyes, and found a corneal epithelial thickness of $49.7~\mu\mbox{m}~(\pm 2.2~\mu\mbox{m})$~\cite{kifi2000};  Li et al used Confocal Microscopy Through Focusing (CMTF) on 7 healthy subjects and found a thickness of $50.6~\mu\mbox{m}~(\pm 3.9~ \mu\mbox{m})$~\cite{lipe97}; and Reinstein et al used Very High Frequency Ultrasound (VHFU) on 110 eyes in 56 subjects who presented for refractive surgery, and they found a thickness of $53.4~\mu\mbox{m}~(\pm 4.6~ \mu\mbox{m})$~\cite{rear2008}. Ladage et al conducted a prospective randomized double-blind study on 246 subjects, assessing the effects of contact lens type and oxygen transmissibility on corneal epithelium in daily wear ~\cite{laya2001}. The study consisted of three treatment groups: high oxygen-transmissible soft lenses (n= 36 subjects), hyper oxygen-transmissible soft lenses (n=135 subjects), and hyper oxygen-transmissible rigid gas permeable lenses (n=75 subjects), in each of which baseline central corneal epithelial thickness was measured using CMTF and found to be $48.55~\mu\mbox{m}~(\pm 3.2~\mu\mbox{m})$, $49.26~ \mu\mbox{m}~(\pm 3.44~\mu\mbox{m})$, and $49.79~\mu\mbox{m}~(\pm 3.25~\mu\mbox{m})$ respectively. There have been a number of studies using optical coherence tomography (OCT)~\cite{husw91,towa05} to estimate central corneal epithelial thickness in humans. Wang et al used OCT on 20 healthy eyes in 40 subjects and found a thickness of $57.8~\mu\mbox{m}~(\pm 1.7~\mu\mbox{m})$ ~\cite{wafo2002}; Feng et al used OCT on 10 healthy subjects and found a thickness of $58.4~\mu\mbox{m}~(\pm 2.5~\mu\mbox{m})$; and Radhakrishnan et al used 1310 nm real-time OCT to make measurements on 5 healthy subjects, and found a thickness of $55~\mu$m ~\cite{raro2001}. Table~(\ref{table:epithelial_thickness}) summarizes the central corneal epithelial thickness for the above mentioned studies. 

\begin{table}[ht] \caption{Thickness of human central corneal epithelium}
\centering
\small{
\begin{tabular}{c|c|c|c}
\hline\hline 
Ref & Method & Sample size & Thickness \\ [0.5ex]
\hline 
~\cite{kifi2000}  & Interferometry & 6 eyes & $49.7~\mu\mbox{m}~(\pm 2.2~\mu\mbox{m})$\\ 
~\cite{raro2001} & Real-time OCT  & 5 subjects & $55~\mu\mbox{m}$\\ 
~\cite{wafo2002} & OCT  & 20 eyes &  $57.8~\mu\mbox{m}~(\pm1.7~\mu\mbox{m})$\\ 
~\cite{fesi2005} & OCT  & 10 subjects & $58.4~\mu\mbox{m}~(\pm 2.5~\mu\mbox{m})$\\ 
~\cite{lipe97} & CMTF & 7 subjects & $50.6~\mu\mbox{m}~(\pm 3.9~\mu\mbox{m})$\\ 
~\cite{laya2001} & CMTF & 36 subjects& $48.55~\mu\mbox{m}~(\pm 3.2~\mu\mbox{m})$\\ 
~\cite{laya2001} & CMTF & 135 subjects & $49.26~\mu\mbox{m}~(\pm 3.44~\mu\mbox{m})$\\ 
~\cite{laya2001} & CMTF & 75 subjects & $49.79~\mu\mbox{m}~(\pm 3.25~\mu\mbox{m})$\\ 
~\cite{rear2008} & VHFU & 56 subjects & $53.4~\mu\mbox{m}~(\pm 4.6~\mu\mbox{m})$\\ 
[1ex]
\hline 
\end{tabular}
}
 \label{table:epithelial_thickness}
\end{table}

\subsubsection{Corneal epithelial refractive index}

Patel et al used a modified hand-held refractometer to measure the refractive index in 10 human eyes in-vivo, and found a value of 1.401 ($\pm$0.005)~\cite{pama95}. There have been few if any other in-vivo estimations of corneal epithelium refractive index in humans. 

\subsection{Bowman's layer}

In humans, Bowman's layer sits posterior to the corneal epithelium and contains randomly arranged collagen fibrils, which on transmission electron microscopy (TEM), appear to be continuous with the more regularly arranged fibrils of the stroma. Fibrils in Bowman's layer were only half to two thirds the diameter of those in the stroma, suggesting that Bowman's layer fibrils may be a condensation of stromal fibrils~\cite{jaje84}.  Using TEM, Hayashi et al found the collagen fibril diameter to be 20 nm~\cite{haos2002} in Bowman's layer and 20-30 nm in the stroma~\cite{haos2002}; and Komai et al found fibril diameters of 25 nm in Bowman's layer and 25-35 nm in the stroma~\cite{kous91}.

\subsubsection{Bowman's layer thickness}
There have been few measurements of human Bowman's layer thickness, most of which have fallen in the $10-17~\mu$m range. In Hayashi et al's TEM studies, measurements were made on an excised conical cornea and found a central thickness of $10~\mu$m. Li et al's CMTF studies on 7 healthy subjects found a thickness of $16.6~\mu\mbox{m}~(\pm 1.1~\mu\mbox{m})$d~\cite{lipe97}; King-Smith et al's interferometry measurements on 6 healthy eyes found a thickness of $14.6~\mu\mbox{m}~(\pm 1.4~\mu\mbox{m})$~\cite{kifi2000}; and  Komai et al's TEM studies on 8 eye-bank eyes found a thickness of 8-12 $\mu$m~\cite{kous91}. Table~(\ref{table:bowman_thickness}) summarizes results from the above sample of work in this area.

\begin{table}[ht] \caption{Central thickness of Bowman's layer}
\centering
\small{
\begin{tabular}{c|c|c|c}
\hline\hline 
Ref & Method & Sample size & Thickness \\ [0.5ex]
\hline 
~\cite{kifi2000}  & Interferometry & 6 eyes & $14.6~\mu\mbox{m}~(\pm 1.4~\mu\mbox{m})$\\ 
~\cite{lipe97} & CMTF & 7 subjects & $16.6~\mu\mbox{m}~(\pm 1.1~\mu\mbox{m})$\\ 
~\cite{haos2002} & TEM & 1 eye & $10~\mu$m\\ 
~\cite{kous91} & TEM & 8 eyes & $8-12~\mu$m\\ 
[1ex]
\hline 
\end{tabular}
}
 \label{table:bowman_thickness}
\end{table}

\subsubsection{Bowman's layer refractive index}
There have been no reported measurements of the refractive index function of Bowman's layer in humans. However, anterior stroma is intermediate between Bowman's layer and posterior stroma, in terms of several structural features such as axial depth, susceptibility to hydration swelling~\cite{mupe2001}, degree of hydration~\cite{leme97}, collagen fibril diameter~\cite{haos2002}, and regularity of fibril arrangement~\cite{frmc95}. It is therefore reasonable to conjecture that the refractive indicex function of the anterior stroma is intermediate between those of Bowman's layer and posterior stroma. This is premised on the notion that the group refractive index of a substance is a function of the refractive indices of the biochemical constituents of that substance. Hence the group refractive index can be estimated by a weighted summation rule such as the Gladstone-Dale relation; or it can be estimated by interpolation, as we do here.

\subsection{Anterior corneal stroma}
The stroma constitutes approximately 90\% of the corneal thickness, and the anterior stroma is the major target for ablation and remodelling in most refractive surgery procedures. The anterior stroma consists of intermediate regularity type I collagen fibrils, 20-35 nm in diameter~\cite{haos2002,kous91}. On scanning electron microscopy (SEM), the collagen fibrils are seen bundled into lamellae which are piled in parallel stacks of about 300 lamellae in the central cornea and about 500 at the limbus~\cite{rama2002}. The anterior stromal lamellae are 0.2–1.2 $\mu$m thick and 0.5–30 $\mu$m wide~\cite{kous91}, and are interconnected at points either via interlamellae crossing collagen fibrils or via anastomosing sublamellae~\cite{rama2002}, consistent with Maurice and Gallagher's earlier findings in the rabbit cornea~\cite{gama77}. The architecture of the anterior stroma confers shape, curvature, rigidity, and hydration resistance to the cornea~\cite{mupe2001, br2001}. 

\subsubsection{Anterior corneal stroma thickness}
The anterior corneal stroma has no conventionally defined thickness or demarcation, but the anterior $100-120~\mu$m of the stroma has been shown to be physiologically distinguishable from more posterior stroma~\cite{mupe2001}. 

\subsubsection{Anterior corneal stroma refractive index}
The refractive index of anterior stroma was observed as $1.38~(\pm0.005)$ by Patel et al, using a bench Abb\'{e} refractometer to make measurements on bare stroma of fresh human corneas~\cite{pama95}. 

\subsection{Posterior corneal stroma}
The posterior stroma differs from anterior stroma in a number of ways. Ultrastructural studies done in rabbit and humans corneas found the posterior stroma to have a greater degree of regularity of collagen fibril arrangement than the anterior stroma~\cite{frmc95}. Unlike anterior lamellae, posterior lamellae may course the entire length of the cornea without branching, and they are typically 1.0 –2.5 $\mu$m thick and 100–200 $\mu$m wide, which is larger in dimension than anterior lamellae~\cite{kous91}. Horizontal branching occurs among both anterior and posterior lamellae, however only anterior lamellae exhibit axial branching~\cite{kous91}. In addition, the posterior stroma has been found to be more susceptible to hydration and swelling than the anterior stroma~\cite{mupe2001}. Though the posterior stroma is largely spared in current refractive surgery procedures, it becomes the cornea's primary source of structural support and bulk following laser ablation of the anterior stroma.

\subsection{Posterior stroma thickness}
The posterior stroma can be reasonably designated as the posterior two-thirds to three-quarters of the stroma. Reinstein et al used VHFU to measure net stromal thickness across the central 10 mm of the cornea in 110 normal eyes, and they found a vertex thickness of $465.4\pm36.9~\mu$m and a thickness of $461.8\pm37.3~\mu$m at the thinnest point~\cite{rear2009}. Erie et al used in-vivo confocal microscopy on 18 eyes of 12 patients presenting for LASIK, and they found a net stromal thickness of $491\pm35~\mu$m pre-operatively~\cite{erpa2002}.

\subsection{Posterior stroma refractive index}
Patel et al used a bench Abb\'{e} refractometer on bare stroma of fresh human cornea, and found a posterior stroma refractive index of $1.373~(\pm0.001)$. This differed from the refractive index of the anterior stroma which they found to be $1.38~(\pm0.005)$~\cite{pama95}.

\subsection{Refractive index distribution}

\begin{figure}[ht]
\begin{center}
\scalebox{.40}
{\includegraphics{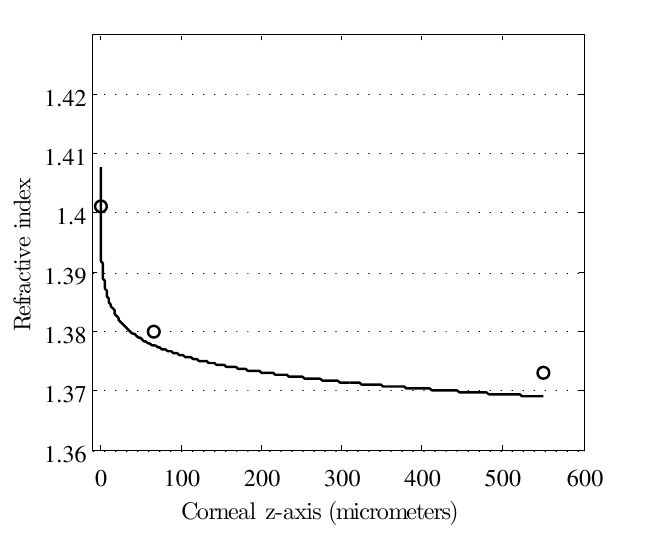}}
\end{center}
\caption{Refractive index sample fit}
\label{fig:RI}
\end{figure}

We fit a curve, shown in Figure~(\ref{fig:RI}), to Patel's experimentally measured data points~\cite{pama95} shown in Table~\ref{table:RI_human_cornea_radial}. The fit curve's expression is given by,

\begin{equation}
n(z) = 1.3952 z^{-0.003}
\label{RI_fcn}
\end{equation}

where $z \geq 0.2$ is the displacement (in $\mu$m) along the optical axis, $n(z)$ is the refractive index at $z$, and we have chosen $z=0~\mu$m as the central corneal epithelial surface, $z=0.2~\mu m$ as the initial observation point of the photon in the corneal epithelium, $z=67~\mu$m as the observation point in the anterior stroma, and $z=550~\mu$m as the observation point in the posterior stroma. These choices are consistent with the literature data.

\begin{table}[h] \caption{Refractive indices in human cornea}
\centering
\small{
\begin{tabular}{c|c|c}
\hline\hline 
Refractometer & Level & Refractive index \\ [0.5ex]
\hline 
modified hand-held  &  epithelium & 1.401 ($\pm$0.005)\\ 
bench Abb\'{e} & anterior stroma  & 1.380 ($\pm$0.005)\\ 
bench Abb\'{e} & posterior stroma & 1.373 ($\pm$0.001)\\ 
[1ex]
\hline 
\end{tabular}
}
 \label{table:RI_human_cornea_radial}
\end{table}

\section{Computation}
\label{sec:compu}
Here we describe the Trajectron algorithm for computing the trajectory of a photon in the human cornea, based on geometric optics and the following two assumptions: (i) non-zero axial GRIN, and (ii) zero meridional GRIN. 

\subsection{The Trajectron Algorithm}

We discretize the optical axis into $N$ segments, with finer segments over Bowman's layer because it has a faster changing refractive index gradient than both the corneal epithelium and stroma. To compute the photon trajectory over the $j$th segment we require: (i) the incident refractive index, $n_j(0)$, (ii) the incident location of the ray, $r_j(0)$, (iii) the refractive index difference quotient ${\hat{\alpha}}_j$, and (iv) the angle ${\theta}_j(0)$ which the incident optical momentum, ${\bM{p}}_j(0)$, makes with the optical axis. In the above notation we have used the domain parametrization, $0 \leq z_j \leq 1$, over each discrete segment $D_j$, such that $f_j$ denotes the restriction of an arbitrary function $f$ over the $j$th segment, $f_j(0)$ denotes the value of $f$ at the start point of the $j$th segment, and $f_j(1)$ denotes the value of $f_j$ at endpoint of the $j$th segment. Domain segment boundary points are related by, $D_j(1) = D_{j+1}(0)$, for every $j$.

(i) We obtain $n_j(0)$ simply by evaluating the fit curve in Equation~(\ref{RI_fcn}). (ii) The incident ray location is given by,

\begin{equation}
 r_j(0) = r_{j-1}(1),
\label{ray_location_update}
\end{equation}

 where $r_{j-1}(1)$, is the ray location at the end of the previous segment and is available from the previous computation.

(iii) The difference quotient is readily computed by,

\begin{equation}
 \widehat{\alpha}_j = \frac{n_j(1) - n_j(0)}{D_j(1)-D_j(0)}.
\label{n_DQ}
\end{equation}

 (iv) We then determine the angle the optical momentum $\bM{p}$ makes with the axial direction by,

\begin{equation}
 \theta_{j+1}\mbox{{\small(0)}} = \mbox{tan}^{-1}\left[{\frac{\partial{r}}{\partial{z}}}\bigg|_{D_j(1)}\right],
\label{theta_eqn}
\end{equation}
 
where,
\begin{equation}
 \frac{\partial{r}}{\partial{z}} =  A\frac{2{\alpha}^2 + (\alpha/G)\left(2{\alpha} n_1(0) + 2{\alpha}^2z\right)}{2{\alpha}^2z + 2{\alpha} n_1(0) + 2\alpha G},
\label{Ray_Gradient_eqn}
\end{equation}

$A$ is a constant defined in Equation~(\ref{A_const}), and $G$ is a function of $z$ defined in Equation~(\ref{G_eqn}).

It follows that,

\begin{equation}
r_{j}(z) = F(j) + r_{j-1}(1), 
\label{ray_updater}
\end{equation}

where $F(j)$ is an evaluation of the closed form solution using the $j$th segment parameters obtained as described above. The Trajectron algorithm is summarized in Table~(\ref{trajectron}).

\begin{table}[h]
 \caption{The Trajectron Algorithm}
\centering
\small{
\begin{tabular}{l}
\hline\hline
\textbf{Input}: $[D_j(0),D_j(1)]$, for  $j=1,...N$\\[0.5ex]
\textbf{Initialize:} $n_j(0)$, $\widehat{\alpha}_j$, $\theta_1(0)$, $r_1(0)=0$, $j=1$  \\ [0.5ex]
$\qquad$ Step 1. F$\big[ n_j(0), \widehat{\alpha}_j, \theta_j(0)\big] + r_{j-1}(1)$ \textbf{\textrightarrow} $r_j$\\ 
$\qquad$ Step 2. $\left.\frac{\partial r}{\partial z}\right|_{D_j(1)}$ \textbf{\textrightarrow} $\gamma $\\
$\qquad$ Step 3. $\mbox{tan}^{-1}(\gamma)$ \textbf{\textrightarrow} $\theta_{j+1}(0)$\\
$\qquad$ Step 4. j \textbf{\textrightarrow} $j+1$\\
$\qquad$ Step 5. loop\\
\hline
\end{tabular}
}
 \label{trajectron}
\end{table}

\section{Results}
\label{sec:rslt}

\begin{figure}[h]
\begin{center}
\scalebox{.45}
{\includegraphics{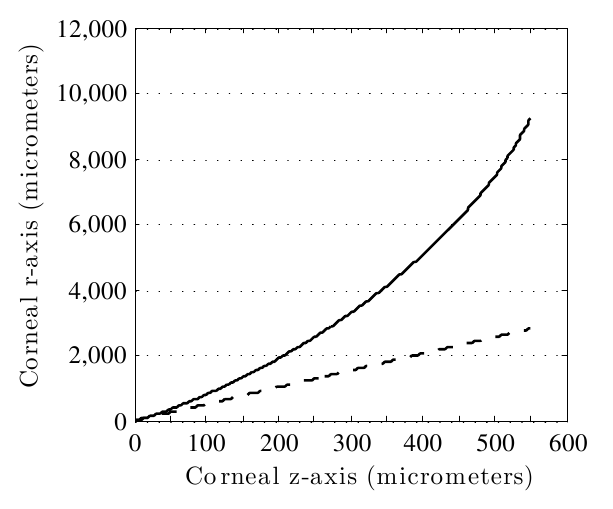}}
\end{center}
\caption{Photon trajectory with ${\theta}_{0.2}$ of 78.97 degrees and ${\theta}_{550}$ of 88.67 degrees (solid line). Also showing initial tangent direction (dashed line)}
\label{fig:bendometer}
\end{figure}

We conducted 40 numerical experiments with $0<{\theta}_{0.2}<\frac{\pi}{2}$, each of which demonstrated bending of the photon trajectory within the human cornea. Results of the first 20 numerical experiments are displayed in Table~(\ref{table:Light bending}). The ${\theta}_{0.2}$ values ranged from 2.21 to 79.05 degrees and resulted in ${\theta}_{550}$ values which ranged from 2.25 to 90 degrees. A photon with a ${\theta}_{0.2}$ of $0$ degrees will proceed along a straight line trajectory. However, for all other ${\theta}_{0.2}$ values, some degree of bending occurs. Figure~(\ref{fig:bendometer}) plots both the photon trajectory (solid line) and the ray initial tangent (dashed line) for a photon with a ${\theta}_{0.2}$ of 78.97 degrees.

\begin{figure}[h]
\begin{center}
\scalebox{.50}
{\includegraphics{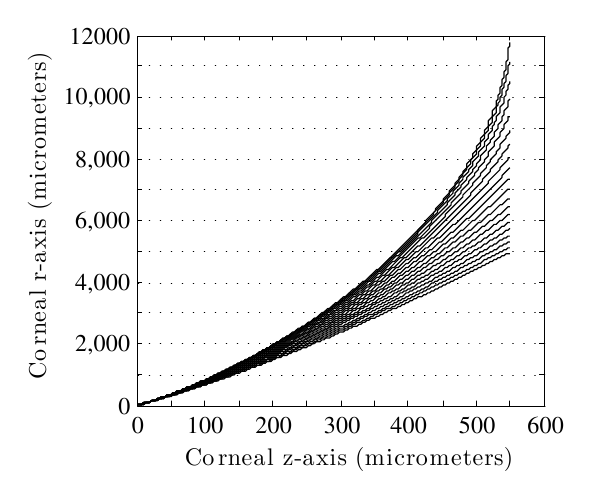}}
\end{center}
\caption{Light broom: 20 photon trajectories with ${\theta}_{0.2}$ ranging from 77.68 to 78.72 degrees and ${\theta}_{550}$ ranging from 85 to 89.75 degrees.}
\label{fig:l_broom_1black}
\end{figure}

 As the ${\theta}_{0.2}$ values approach 90 degrees, the bending of the photon trajectory increases, and becomes increasingly and exquisitely dependent on angle. The difference, $\delta$, between the initial tangent line and the photon trajectory is a measure of the extent of bending. We use ${\delta}_z$ to denote the displacement along the r-axis at position $z~\mu$m due to light bending. Table~(\ref{table:Light bending}) shows the ${\delta}_z$ values at $70\mu$m, $200\mu$m, $400\mu$m, and $550\mu$m. In the next 20 numerical experiments, we computed photon trajectories with ${\theta}_{0.2}$ values clustered closer to 90 degrees. Specifically, the ${\theta}_{0.2}$ values ranged from 77.68 to 78.72 degrees and resulted in ${\theta}_{550}$ values ranging from 85 degrees to 89.75 degrees. The photon trajectories were noted to bend away from the optical axis in a concave upwards direction as shown in Figure~(\ref{fig:l_broom_1black}).

\begin{table}[h] \caption{Photon trajectories in the human cornea}
\centering
\small{
\begin{tabular}{c|c|c|c|c|c}
\hline\hline 
${\theta}_{0.2}$  & ${\theta}_{550}$  & $\delta_{70}$ & $\delta_{200}$  & $\delta_{400}$ & $\delta_{550}$\\ [0.5ex]
(in degrees) & (in degrees) & (in $\mu$m) & (in $\mu$m) & (in $\mu$m) & (in $\mu$m)\\
\hline 
79.05 & 90.00 & 58.24 & 299.79 & 916.87 & 9728.70\\ 
78.84 & 87.80 & 53.65 & 270.30 & 801.77 & 4983.10\\ 
78.25 & 85.71 & 43.60 & 210.14 & 590.23 & 2839.90\\ 
76.36 & 81.82 & 24.82 & 110.96 & 289.03 & 1144.10\\ 
74.00 & 78.26 & 14.20 & 060.90 & 153.21 & 0567.50\\ 
71.50 & 75.00 & 08.75 & 036.71 & 090.79 & 0326.45\\ 
66.63 & 69.23 & 04.12 & 016.94 & 041.25 & 0144.63\\ 
62.20 & 64.29 & 02.39 & 009.71 & 023.49 & 0081.53\\ 
58.24 & 60.00 & 01.58 & 006.38 & 015.38 & 0053.10\\ 
54.72 & 56.25 & 01.14 & 004.59 & 011.03 & 0037.98\\ 
50.14 & 51.43 & 00.78 & 003.13 & 007.51 & 0025.76\\ 
43.97 & 45.00 & 00.49 & 001.98 & 004.74 & 0016.25\\ 
35.25 & 36.00 & 00.28 & 001.12 & 002.68 & 0009.15\\ 
29.40 & 30.00 & 00.19 & 000.78 & 001.89 & 0006.38\\ 
22.07 & 22.50 & 00.12 & 000.50 & 001.18 & 0004.04\\ 
17.66 & 18.00 & 00.09 & 000.37 & 000.88 & 0003.00\\ 
11.78 & 12.00 & 00.06 & 000.23 & 000.54 & 0001.86\\ 
08.83 & 09.00 & 00.04 & 000.17 & 000.40 & 0001.36\\ 
04.42 & 04.50 & 00.02 & 000.08 & 000.19 & 0000.66\\ 
02.21 & 02.25 & 00.01 & 000.04 & 000.10 & 0000.33\\ 
[1ex]
\hline 
\end{tabular}
}
 \label{table:Light bending}
\end{table}

\section{Discussion}
\label{sec:disc}
Until recently, schematic eye models assumed a constant or average corneal refractive index~\cite{atsm2000,sm1995,ma57,ma1994}. Since emerging evidence of corneal GRIN nature, a few newer models have respected the GRIN. But like their older counterparts, these newer models are based on paraxial approximation and ray transfer matrix analysis~\cite{fldi2009, peba2005, ba2006}. The first-order paraxial approximation of the wave equation in shortwavelength limit, is one whose solutions are rays propagating along straight lines. And this linearity allows an optical system to be modelled using ray-transfer matrix analysis. The fitness of these ray-transfer matrix analyses depend on the magnitude of high frequency coefficients in the fourier transform of the refractive index function. In the case of highly varying refractive index gradient and consequently highly varying ray spatial gradient, linear approximations have limited validity. And a prohibitively large number of optical elements or ray transfer matrices may be required to obtain a good approximation of the ray path. The transition zone between the corneal epithelium and Bowman's layer is an example of such a steep gradient, where finer linearization granularity is required. There is not a general notion of when linear approximation approaches may begin to lose validity, as the small angle approximation, sin$(\theta)\approx\theta$, has no defined boundary. For this reason the angular domain over which the paraxial approximation is clinically applicable is not clear. Yet clinically, refractive error measurements and correction are based on paraxial approximation. Results of our numerical experiments shown in Figure~(\ref{fig:l_broom_1black}) may help us gain some traction on this issue. We see an increase in non-linearity with ${\theta}_{0.2}$. This increase in non-linearity is correlated with a loss of paraxial approximation validity. Also from Figure~(\ref{fig:l_broom_1black}), we see that linear-like form is preserved over a much wider angle than only those for which sin$(\theta)\approx\theta$. This suggests that the angular domain over which the paraxial approximation is clinically applicable is more broad than suggested by the small angle approximation alone. For instance, for a ${\theta}_{0.2}$ of 50 degrees, the $\delta_{550}$ is $25~\mu$m, which is a relatively small displacement. In addition to exhibiting such preservation of linear-like form over broad angle, the paraxial approximation is a potent simplification device in theory and practice. For these reasons, it has enjoyed much success and longevity in the clinic. However, its domain of validity is limited. For instance, for ${\theta}_{0.2} >$ 75 degrees, the corresponding  $\delta_{550}$ exceeds $0.5$ mm and increases rapidly and non-linearly with  ${\theta}_{0.2} $; such that by  ${\theta}_{0.2}=78.84$ degrees, the $\delta_{550}$ has reached 5 mm, which is almost the entire radius of the cornea. This is clearly non-paraxial territory over which the paraxial approximation is invalid. Clinically, this zone is at the fringes of peripheral vision. Loss of vision here is not likely to interfere significantly with daily life, and may even go unnoticed. However, such subtle peripheral vision loss is often the first sign of an insidious optic neuropathy such as primary open angle glaucoma.

In the numerical experiments presented here, the scope of the photon trajectories studied are exclusively intra-corneal, with $z \in [0.2,550]~\mu$m. Based on these purely intra-corneal photon trajectories, we have made inference involving the visual field. This is justified by a reasonable assumption of continuity. We have assumed that there exists a correspondence between the intra-corneal angle, $\theta$, and the $\theta$ of the originating ray source in the visual field. In other words, relatively wide angles in the field give rise to relatively wide angles within the cornea, and relatively narrow angles in the field give rise to relatively narrow angles within the cornea. However, while we have relied on this angular correspondence to make inference on field source, it is important to note that what we have shown in this paper is that this correspondence is non-linear. And more specifically, we have shown that wide intra-corneal angles give rise to wider intra-corneal angles, while narrow intra-corneal angles remain relatively constant.

Limitations of the current study include the assumption of gradient zero in the meridional directions. Vasudevan et al showed naso-temporal gradients in the human corneal epithelium ~\cite{vasi08}. Using a portable Abb\'{e} refractometer mounted on a slit-lamp to make measurements on 10 human subjects, they found corneal vertex, nasal, and temporal refractive indices of $1.397~(\pm 0.001)$, $1.394~(\pm 0.001)$, and $1.394~(\pm 0.001)$ respectively. These gradients are significantly less than those found by Patel~\cite{pama95} in the axial direction, and hence were assumed to be zero. The naso-temporal gradients were in the direction of the corneal vertex, and their incorporation into the model can be expected to attenuate the concavity of the photon trajectory to some degree. However, the overall non-linearity and angular-dependence behavior shown here will be unchanged. A second limitation of the current work is the non-uniform specimen preparation and measurement method of the axial refractive index. The epithelial refractive index was measured in-vivo using a modified hand-held refractometer, while the stroma was prepared as bare fresh cornea and measured in-vitro using a bench-top Abb\'{e} refractometer~\cite{pama95}. A third limitation of the current study is the sparsity of refractive index data points currently available to us in the literature. This sparsity is due to the technological difficulty of making fine high precision refractive index measurements in the human cornea. Most of this challenging work has been done by Patel and by Vasudevan. Both Patel's axial measurements and Vasudevan's naso-temporal measurements contained only three data points each, which points towards the challenging nature of the work. Unfortunately, mathematical models based on such sparse data may be susceptible to the aliasing problem. For instance, in the current study we only had one data point available for the refractive index gradient in the corneal epithelium. However, the corneal epithelium is $55~\mu$m thick on average with 5-7 cell layers and contains histophysiological gradients. The cells are older, flatter, and closer to apoptosis anteriorly than posteriorly The epithelium is bathed in the tear film mucinous phase anteriorly and continuous with bowman's layer posteriorly. These histophysiological gradients suggest that the corneal epithelium contains non-zero refractive index gradients. If a gradient does exists, its direction and form beg investigation. More generally, finer sampling granularity of the refractive index is needed in the human cornea in each of the three principal axis directions. In-spite of the above limitations, the current study has provided a clear first demonstration of non-paraxial intra-corneal light-bending, and highlighted potential areas for further experimental and theoretical collaboration in addressing the global disease burden of refractive error.

\section{Conclusion}
\label{sec:conclu}
In this paper, we have provided the first quantitative demonstration of non-paraxial light-bending within the human cornea. We have also introduced the Trajectron algorithm for computing photon trajectories in the human cornea. And we have initiated steps towards describing the angular domain over which the paraxial approximation is clinically applicable. These demonstrations have significant implications in the development of refractive surgery algorithms.

\section*{Acknowledgement}
I would like to express gratitude to my wife Lisa for her moral support. I would like to thank my friend Peter Blair for reading this manuscript and providing helpful suggestions on its technical presentation.

\providecommand{\bysame}{\leavevmode\hbox to3em{\hrulefill}\thinspace}
\providecommand{\MR}{\relax\ifhmode\unskip\space\fi MR }
\providecommand{\MRhref}[2]{%
  \href{http://www.ams.org/mathscinet-getitem?mr=#1}{#2}
}
\providecommand{\href}[2]{#2}

\newpage
\section*{Author Biography}

{\small{
Dr. Stephen G. Odaibo is an ophthalmology resident at Howard University in Washington DC. He attended Duke University School of Medicine where he was awarded the Barrie Hurwitz Award for Excellence in Clinical Neurology. He is an alumni of the Mathematics Fast Track Program at the University of Alabama at Birmingham (UAB) where he was awarded the International Scholar Award for Academic Excellence. Dr. Odaibo received a Bachelors degree with Honors in Mathematics from UAB in 2001, an M.S. degree in Mathematics from UAB in 2002, and an M.S. degree in Computer Science from Duke University in 2009. He received an M.D. degree from Duke University School of Medicine in 2010 and completed a housemanship in Internal Medicine at Duke University Hospital in 2011. Dr. Odaibo's research interests are at the intersection of Mathematics, Computer Science, Physics, and Biomedical science, with a special focus on the interaction of light with biological systems. He is happily married to Dr. Lisa Marie Odaibo who is completing a pediatrics residency at Georgetown University.}}
\end{document}